# Structure and magnetic properties of a family of two-leg spin ladder compounds $Ba_2RE_2Ge_4O_{13}$ (RE = Pr, Nd, and Gd-Ho) with strong rung interaction

Jin Zhou,[a] Andi Liu,[a,b] Fangyuan Song,[a] Langsheng Ling,[c,*] Jingxin Li,[c] Wei Tong,[c] Zhengcai Xia,[a] Gaoshang Gong,[d] Yongqiang Wang,[d] Jinkui Zhao,[b,e] Hanjie Guo,[b,*] Zhaoming Tian [a,*]

[a] School of Physics and Wuhan National High Magnetic Field Center, Huazhong University of Science and Technology, Wuhan, Hubei, 430074, China

[b] Songshan Lake Materials Laboratory, Dongguan, Guangdong, 523808, China

[c] Anhui Key Laboratory of Low-energy Quantum Materials and Devices, High Magnetic Field Laboratory, HFIPS, Chinese Academy of Sciences, Hefei, Anhui, 230031, China

[d] Zhengzhou University of Light Industry, School of Electronics & Information, Zhengzhou, Henan, 450002, China

[e] School of Physical Sciences, Great Bay University, Dongguan, Guangdong, 523000, China

## ABSTRACT:

Compared to the intensive investigation on the 3$d$ transition-metal (TM)-based spin ladder compounds, less attention has been paid to the ones constructed by the rare-earth (RE) ions. Herein, we report a family of RE-based spin ladder compounds $Ba_2RE_2Ge_4O_{13}$ (RE = Pr, Nd, Gd-Ho) crystallized into the monoclinic structure with the space group $C2/c$. The RE ions are arranged on a two-leg spin ladder motif along the $b$-axis, where the rung and leg exchange interactions are bridged via the RE-O-RE pathways and RE-O-Ge-O-RE routes, respectively. Moreover, the much shorter rung distance in the $RE_2O_{12}$ dimer units than the leg distance suggests $Ba_2RE_2Ge_4O_{13}$ to be a strong-rung spin ladder system. All the synthesized $Ba_2RE_2Ge_4O_{13}$ (RE = Pr, Nd, Gd-Ho) compounds exhibit the dominant antiferromagnetic (AFM) interactions and absence of magnetic order down to 1.8 K. Among the family members, $Ba_2Dy_2Ge_4O_{13}$ can be described by $J_{eff}$ = 1/2 Kramers doublet states, the low temperature specific heat indicates the coexistence of spin dimerized state with broad maximum at ~ 2.4 K and long-range AFM order with $T_N$ = 0.81 K. This family of $Ba_2RE_2Ge_4O_{13}$ compounds thereby provides a rare platform to investigate the novel spin ladder physics constructed by 4$f$ electrons.

## 1. INTRODUCTION

The spin ladders, as one archetype of low-dimensional quantum magnets, play an important role in studying the crossover from one-dimensional (1D) spin chain to the two-dimensional (2D) square-lattice magnets.[1,2] For the 1D spin chain system, significant progress has taken place on clarifying its magnetic ground states and related physical phenomena,[3,4] as one typical example, the uniform Heisenberg antiferromagnetic (AFM) spin chain with half-integer spin ($S$ = 1/2) has been established to show no long-range magnetic order with gapless excitations.[5] The 2D square-lattice magnets always form the long-range AFM order even in the extreme quantum limit of $S$ = 1/2.[6,7] By contrast, spin ladders can display significantly different magnetic ground states in terms of the number (n) of legs to be even or odd.[2,8-11] The odd-leg ladders, such as the three-leg $Sr_2Cu_3O_5$,[11,12] show the gapless spin excitations with power-law decaying spin correlations similar to those of 1D spin chain. On the other side, the even-leg ladders form the quantum disordered singlet state with exponentially decaying spin correlations due to a finite spin gap,[13,14] as experimentally verified in two-leg $SrCu_2O_3$.[15] To date, numerous spin ladder compounds have been synthesized and magnetically characterized, which exhibit diverse collective quantum phenomena including the field-induced Luttinger spin liquid,[16] multi-triplon excitations,[17,18] Wigner crystallization[19] and Bose-Einstein condensation of magnons.[20-22]

One of the most extensively investigated spin-ladder system is the two-leg $S$ = 1/2 ladder materials,[2,23] whose ground states are nontrivial and depend critically on the relative signs and strengths of the exchange couplings of rungs ($J_{rung}$) and legs ($J_{leg}$) of the ladder. The S = 1/2 two-leg spin ladders, as changing the ratio of $J_{rung}/J_{leg}$, can be categorized into the strong-rung, strong-leg and nearly isotropic spin ladders. At various coupling limits, their magnetic ground states are quite different showing rich magnetic phase diagrams. In the case of strong-rung ($J_{rung}/J_{leg} \gg 1$) coupling, it can be viewed as an isolated spin dimer with a singlet ground state, as realized in many real materials such as $(C_5H_{12}N)_2CuBr_4$,[14,24] $(VO)_2P_2O_7$[25] and $Rb_3Ni_2(NO_3)_7$,[26] etc. In the weak rung-coupling regime, exotic magnetic states such as columnar dimer phase and staggered dimer phase have also been proposed,[27,28] as experimentally realized in $(C_7H_{10}N)_2CuBr_4$.[29] Correspondingly, the compounds can be approximately described by a pair of weakly coupled uniform spin chains, and the bound magnons (S = 1) excitations can emerge in contrast to strong-rung ladder with singlet-triplet transitions.[30,31] Moreover, in the two-leg spin ladders, the introduction of lattice geometric frustration and single-ion anisotropy can further enrich the exotic magnetic phases, which have been theoretically predicted but not well studied in experiments. Thus, the continuous investigations on new two-leg spin ladder materials are important for enriching the exotic magnetic states and related physical phenomena.

Recently, the search for quantum spin liquid (QSL) state has triggered the new materials' exploration on low-dimensional frustrated magnets consisting of rare-earth (RE) ions.[32-34] Due to the intrinsic strong spin-orbit coupling (SOC), crystalline electric field (CEF) effects and diverse single-ion anisotropy of RE ions, the RE-based compounds offer an alternative platform for realizing the exotic magnetic states beyond 3$d$ transition-metal (TM) -based ones. Indeed, the materials with various lattice geometry motifs have been discovered, including the 2D triangular-lattice magnets,[33,35] Kagome magnets[36,37] and honeycomb lattice compounds.[38,39] While, little attention has been devoted to the RE-based spin ladder systems. Compared to the large exchange couplings with energy scale of hundreds of Kelvin in 3$d$-based spin ladders, the small exchange interactions of several Kelvin between the magnetic moments of RE$^{3+}$ ions allow for the tunability by low magnetic fields accessible in the laboratory settings. This advantage makes them attractive for precise control on the rung/leg exchange couplings, spin gap, and exotic magnetic ground states. Unfortunately, only the metallic REB$_{50}$(RE = Tb, Dy, Ho, Er, Tm) borides with a structural spin-ladder model have been magnetically characterized.[40,41] Even magnetic behaviors of some family members REB$_{50}$ have been discussed in terms of the formation of coupled RE-dimers,[41,42] the existence of Ruderman-Kittel-Kasuya-Yosida (RKKY) interactions between the RE ions complicate the underlying physics of exotic magnetism. Therefore, the insulating compounds are crucial to uncover the intrinsic spin ladder physics fully dictated by RE$^{3+}$ ions, that requires the discovery of RE-based spin ladder materials but remain largely unexplored.

The RE-based Ba$_2$RE$_2$Ge$_4$O$_{13}$ (RE = Pr, Nd, Gd-Dy) tetragermanates have been previously studied for their interesting optical and luminescence properties.[43] Despite the well-established crystal structure of monoclinic structure with space group C2/c (Z = 4),[44] the lattice geometry of RE$^{3+}$ ions and magnetic behaviors have not yet been unveiled. Here, we report the synthesis and magnetic properties on this family of Ba$_2$RE$_2$Ge$_4$O$_{13}$ (RE = Pr, Nd, Gd-Ho) compounds. The structural analyses reveal that RE ions are structurally arranged on the two-leg spin ladder running along the $b$-axis, where RE$^{3+}$ ions are linked through the RE-O-RE pathways and the RE-O-Ge-O-RE routes along the rung and leg directions, respectively. All synthesized compounds exhibit the dominant AFM interactions at low temperatures and absence of long-range magnetic order down to 1.8 K. Among them, Ba$_2$Dy$_2$Ge$_4$O$_{13}$ can be described by the strong-rung spin ladder model with an effective $J_{eff}$ = 1/2 moment, and the specific heat data indicates the coexistence of spin dimerized state with broad maximum at ~ 2.4 K and long-range AFM order with $T_N$ ~ 0.81 K.

## 2. EXPERIMENTAL SECTION

### 2.1. Material Synthesis.

All the Ba$_2$RE$_2$Ge$_4$O$_{13}$ (RE = Pr, Nd, Gd-Ho) polycrystalline samples were synthesized by the conventional high temperature solid-state reaction method using BaCO$_3$ (99.9%, Aladdin), GeO$_2$ (99.9%, Aladdin), and RE oxides (RE = Pr, Nd, Gd-Ho; 99.99%, Aladdin) as starting materials. Prior to use, the $RE_2O_3$ ( RE = Pr, Nd, Eu) powders were heated at 900°C for 12 hours to remove the crystal water. Stoichiometric amounts of the starting materials with a molar ratio of Ba/RE/Ge = 1:1:2 were accurately weighed and thoroughly mixed, then the mixtures of the raw materials were thoroughly ground for 2 hours and pre-reacted at 1000°C for 24 hours in air. Subsequently, the obtained powders were ground again, pressed into pellets and sintered at 1100°C for 2 days with intermediate grindings to get the polycrystalline samples. For RE = Dy and Ho, a higher reaction temperature of 1200°C is required to obtain better crystallization of polycrystalline samples. All the synthesized samples were air-stable.

**2.2. Structure Characterization.**

The $Ba_2RE_2Ge_4O_{13}$ (RE = Pr, Nd, Gd-Ho) polycrystalline samples were characterized using a PAN Analytical X'Pert Pro MPD diffractometer with Cu Kα radiation (λ = 1.5418 Å). The X-ray diffraction (XRD) peaks were recorded from 10 to 70° in a $2\theta$ range with 0.02° step. The Rietveld refinements of XRD data at room temperature were performed using the general structural analysis system (GSAS) program.[45]

**2.3. Physical Property Measurements.**

Temperature (*T*) dependence of magnetic susceptibility and isothermal field ($\mu_0H$)-dependent magnetizations of $Ba_2RE_2Ge_4O_{13}$ compounds were carried out using a superconducting quantum interference device (SQUID, Quantum Design) equipped in Magnetic Properties Measurement System (MPMS3). The direct current (*dc*) magnetic susceptibility in a temperature range from 1.8 K to 300 K was measured using the field-cooling mode under different magnetic fields, the alternating current (*ac*) magnetic susceptibility was also measured at frequencies of 9 to 997 Hz in a temperature range from 1.8 K to 15 K. The isothermal magnetization measurements *M* ($\mu_0H$) were performed in field range from -7 T to 7 T. At temperatures below 1.8 K, magnetic characterizations were carried out by the MPMS3 equipped with a He3 option down to 0.5 K. The X-band (*f* = 9.4 GHz) electron spin resonance (ESR) measurements were carried out using a Bruker spectrometer at the High Magnetic Field Laboratory of the Chinese Academy of Science. For $Ba_2Dy_2Ge_4O_{13}$ sample, the specific heat $C_p(T)$ measurements were performed using the physical properties measurement system (PPMS, Quantum Design) at different magnetic fields. The ultra-low temperature specific heat with temperature down to 100 mK was measured in PPMS using the heat capacity option equipped with a dilution refrigeration system.

## 3. RESULTS AND DISCUSSION

**3.1. Description of Crystal Structure.**

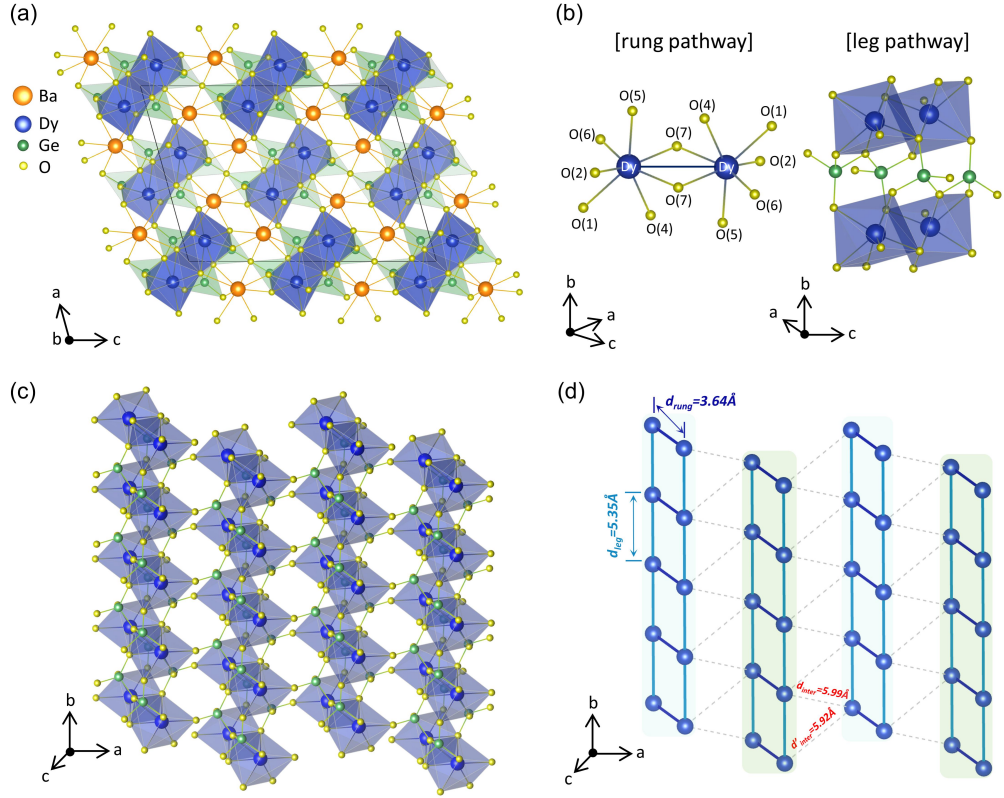

**Figure 1**. (a) The schematic crystal structure of $Ba_2Dy_2Ge_4O_{13}$. The orange, blue, green, and yellow balls denote the Ba, Dy, Ge and O atoms, respectively. (b) The connections of $Dy^{3+}$ ions along the rung- and leg- pathway. (c) The two-leg spin ladders constructed by $Dy^{3+}$ ions running along the *b*-axis, the neighboring $Dy^{3+}$ ions are connected by nonmagnetic $GeO_4$ tetrahedra. (d) The lattice geometry of spin ladder of $Dy^{3+}$ ions and its stacking sequences extended along the *a*-axis. The nearest-neighbor ladders are interconnected in a zigzag fashion, the rung- and leg- distances are denoted by $d_{rung}$ and $d_{leg}$, and the distances of nearest-neighbor ladders are denoted by $d_{inter}$ and $d'_{inter}$.

The synthesized $Ba_2RE_2Ge_4O_{13}$ (RE = Pr, Nd, Gd-Ho) polycrystalline samples are crystallized into the monoclinic system with space group *C*2/*c* (No.15, Z = 4), similar to the previous report.[43] From the structure refinements of powder XRD spectra, the obtained bond distances, bond angles, and intra-ladder and inter-ladder RE-RE distances for $Ba_2RE_2Ge_4O_{13}$(RE = Pr, Nd, Gd-Ho) polycrystals are listed in Table 1. Since all family members are isostructural, as a representative, the crystal structure of $Ba_2Dy_2Ge_4O_{13}$ is schematically shown in Figure 1a. The unit cell consists of 11 crystallographic sites, including one Dy atom (Wyckoff site 8*f*), one Ba atom (Wyckoff site 8*f*), two Ge atoms (Wyckoff site 8*f*), and seven O atoms (Wyckoff site 8*f* and 4*e*), respectively. Each magnetic $Dy^{3+}$ ion is coordinated into the distorted $DyO_7$ polyhedral environment, and then two neighboring $DyO_7$ polyhedra are linked through edge-sharing to form a structural dimer unit of $Dy_2O_{12}$, as shown in Figure 1b. The $Ge^{4+}$ ions are coordinated into $GeO_4$ tetrahedron, and four $GeO_4$ tetrahedra are connected into a zigzag $Ge_4O_{13}$ tetramer through the bridging oxygen atoms. Along the *b*-axis, the neighboring $Dy_2O_{12}$ dimers are linked by four $GeO_4$ tetrahedra, their linkages are realized through co-sharing edge with tetrahedral $Ge1O_4$ units and co-shared angles with $Ge2O_4$ tetrahedra. In the crystal structure, the $Ba^{2+}$ ions are located in the voids between the magnetic $Dy^{3+}$ layers with the $BaO_{10}$ coordination environment. As listed in Table 1, the Dy-O bond distances are ranging from 2.10229(2) to 2.57703(3) Å in the $Dy_2O_{12}$ dimers with Dy-O7-Dy bond angle of 96.695(1)°. The Ba-O bond lengths range from 2.7050(4) to 3.2981(4) Å in the $BaO_{10}$ polyhedra. The distorted $Ge1O_4/Ge2O_4$ tetrahedra have four different Ge1/Ge2-O bond distances ranging from 1.61730(2) to 1.89809(2) Å, and the Ge1-O3-Ge1 and Ge1-O4-Ge2 bond angles are 133.630(1)° and 113.143(1)°, respectively.

As shown in Figure 1a-c, magnetic $Dy^{3+}$ ions in $Ba_2Dy_2Ge_4O_{13}$ are structurally arranged on a two-leg spin ladder geometry extended along the *b*-axis. The rungs of this ladder are formed by two Dy-O-Dy exchange paths in the $Dy_2O_{12}$ dimers (see Figure 1b), while the legs involve a longer interaction path through the super-superexchange Dy-O-Ge-O-Dy routes along the *b*-axis. Further checking the space separation between the $Dy^{3+}$ ions, we can find there is a large difference along the rung- and leg- directions with $d_{rung}$~3.6442(5) Å and $d_{leg}$~5.34101(9) Å, respectively. Within the *ac*-plane, these infinite quasi-1D two-leg ladders are well separated by the nonmagnetic $GeO_4/BaO_{10}$ polyhedra, the adjacent ladders are linked through the $GeO_4$ tetrahedral units with the Dy-O-Ge-O-Dy and Dy-O-Ge-O-Ge-O-Dy super-superexchange pathways along the *a*- and *c*- axes (see the details in Figure S1 in Supporting Information). Moreover, the nearest-neighbor (NN) inter-ladder distances $d_{inter,a}$ = 5.9954(7) Å and $d'_{inter,a}$ = 5.9168(7) Å along the *a*-axis and $d_{inter,c}$ = 6.6107(9) Å along the *c*-axis are larger than $d_{leg}$. In this case, the inter-ladder exchange couplings ($J_{inter,a}$, $J'_{inter,a}$) should be smaller than the leg exchange interactions ($J_{leg}$) and rung exchange interaction ($J_{rung}$), then $J_{rung}$ and $J_{leg}$ are the NN and next-nearest-neighbor (NNN) exchange interactions. As depicted in Figure 1d, the neighboring ladders are coupled by the zigzag type fashion, the slight difference between $d_{inter,a}$ and $d'_{inter,a}$ is related to the variations of the Ge-O3 and Ge-O4 bond distances within the distorted $GeO_4$ tetrahedra. Here, the much shorter $d_{rung}$ than $d_{leg}$ and differences in superexchange pathways support the $J_{rung}$ is much larger than $J_{leg}$. Thus, $Ba_2RE_2Ge_4O_{13}$ are expected to realize a strong-rung spin ladder system.[2,46] In case of the dipolar energy scaling as $1/r^3$,[47] the NN dipole-dipole interactions within the rungs are ~3.2 times larger than the ones along the legs.

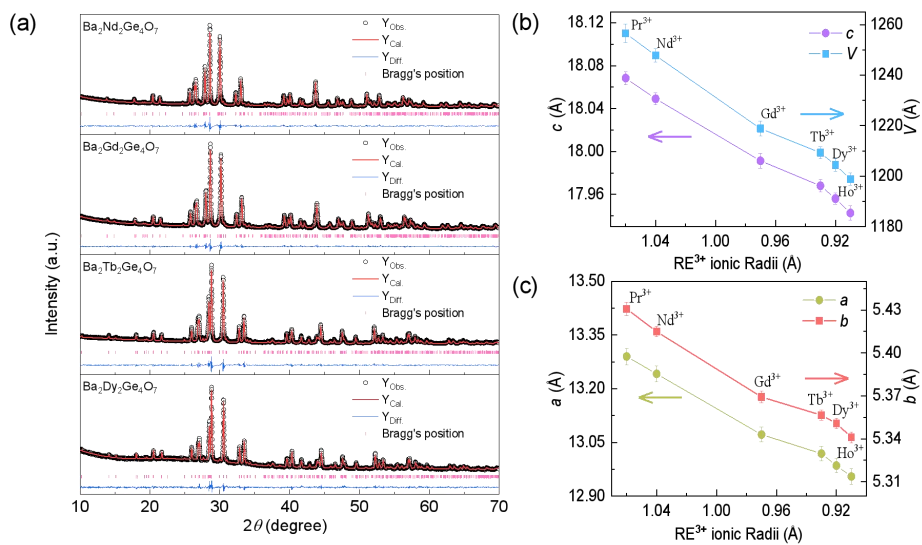

**Figure 2**. (a) Room-temperature powder X-ray diffraction (PXRD) patterns of $Ba_2RE_2Ge_4O_{13}$(RE = Nd, Gd, Tb and Dy) compounds. The black open circles and red lines denote the experimental data and the calculated spectra, blue lines show their differences, and the pink sticks denote the Bragg reflection positions. (b, c) The variation of lattice parameters (*a*, *b*, *c*) and volume of the unit cell as function of RE ionic radii of $Ba_2RE_2Ge_4O_{13}$.

Room-temperature powder XRD patterns for the selected $Ba_2RE_2Ge_4O_{13}$ (RE = Nd, Gd, Tb, Dy) samples are shown in Figure 2a and Figure S2. The calculated XRD patterns yield the good fits to the experimental spectra, no impurity phases are detected in the XRD spectra. The obtained crystallographic data and refinement parameters of the synthesized $Ba_2RE_2Ge_4O_{13}$ samples are listed in Table 1 and Table S1. The $R_{wp}$, $R_p$, and $\chi^2$ factors converge to the values below 5%, indicating the reliability of refined crystal structure. As shown in Figure 2b,c, the lattice parameters (*a*, *b*, *c*) and unit cell volume (*V*) follow a monotonic decrease with the $RE^{3+}$ ionic radii, which can be related to the reduction in the radius of the $RE^{3+}$ ion from $Pr^{3+}$ to $Ho^{3+}$. Using the obtained lattice parameters, the distances of rungs and legs of ladders are calculated as shown in Table 1. The $d_{rung}$ decreases from 3.78977(6) Å (RE = Pr) to 3.66821(7) Å (RE = Ho), and $d_{leg}$ is reduced from 5.43076(7) Å (RE = Pr) to 5.34101(9) Å (RE = Ho). It is also noted that the structure refinements reveal no significant antisite occupations between magnetic $RE^{3+}$ ions and nonmagnetic $Ge^{4+}/Ba^{2+}$ ions possibly due to the large difference of ionic radii and coordination numbers of RE and Ge/Ba cations. Thus, the $RE^{3+}$ ions in $Ba_2RE_2Ge_4O_{13}$ form the fully ordered spin ladder motif with weak antisite disorder effect.

**Table 1.** The selected bond distances, bond angles of $Ba_2RE_2Ge_4O_{13}$ (RE = Pr, Nd, Gd-Ho) polycrystals

| RE | Pr | Nd | Gd | Tb | Dy | Ho |
|---|---|---|---|---|---|---|
| Crystal system | Monoclinic | Monoclinic | Monoclinic | Monoclinic | Monoclinic | Monoclinic |
| Space group | C2/c | C2/c | C2/c | C2/c | C2/c | C2/c |
| Intra-ladder RE-RE (Å) | 3.78977(6) | 3.77351(8) | 3.77846(11) | 3.70312(7) | 3.6442(5) | 3.66821(7) |
| | 5.43076(7) | 5.41491(11) | 5.36920(17) | 5.35654(9) | 5.3507(11) | 5.34101(9) |
| Inter-ladder RE-RE (Å) | 5.93904(6) | 5.93461(8) | 5.85642(13) | 5.89736(7) | 5.9168(7) | 5.82696(7) |
| | 6.06114(6) | 6.04029(8) | 5.96567(13) | 6.01433(7) | 5.9954(7) | 5.99804(7) |
| RE-O1 (Å) | 2.53706(2) | 2.45866(4) | 2.39201(6) | 2.58076(3) | 2.36605(3) | 2.32687(3) |
| RE-O2 (Å) | 2.44523(2) | 2.48194(4) | 2.45265(6) | 2.46184(4) | 2.41756(3) | 2.32423(3) |
| RE-O4 (Å) | 2.55749(2) | 2.57571(3) | 2.55881(5) | 2.55316(2) | 2.57703(3) | 2.57809(3) |
| RE-O5 (Å) | 2.25246(2) | 2.35537(4) | 2.31727(6) | 2.10599(3) | 2.10229(2) | 2.03808(3) |
| RE-O6 (Å) | 2.44536(3) | 2.39664(5) | 2.31742(7) | 2.45906(3) | 2.4704(5) | 2.24858(3) |
| RE-O7 (Å) | 2.63648(3) | 2.60464(5) | 2.47360(7) | 2.67728(3) | 2.5633(4) | 2.60881(4) |
| RE-O7 (Å) | 2.40285(3) | 2.37058(4) | 2.30970(7) | 2.30430(4) | 2.3088(4) | 2.50822(3) |
| RE-O7-RE (°) | 97.428(1) | 98.548(1) | 104.305(2) | 95.744(1) | 96.695(1) | 91.574(1) |
| Ge1-O1 (Å) | 1.67338(2) | 1.74659(3) | 1.71870(5) | 1.54885(2) | 1.74143(3) | 1.55558(2) |
| Ge1-O2 (Å) | 1.63439(1) | 1.68247(2) | 1.70791(4) | 1.69205(2) | 1.68981(2) | 2.00741(2) |
| Ge1-O3 (Å) | 1.78081(2) | 1.77643(3) | 1.76747(4) | 1.75965(2) | 1.84140(2) | 1.78409(2) |
| Ge1-O4 (Å) | 1.67200(2) | 1.71671(3) | 1.84080(5) | 1.78607(3) | 1.89809(2) | 1.95417(3) |
| Ge2-O4 (Å) | 1.95199(2) | 1.88558(3) | 1.80289(4) | 1.85863(2) | 1.74166(1) | 1.74766(2) |
| Ge2-O5 (Å) | 1.74196(2) | 1.68601(3) | 1.70599(5) | 1.73637(2) | 1.64889(2) | 1.82112(2) |
| Ge2-O6 (Å) | 1.69590(2) | 1.72642(3) | 1.70079(4) | 1.59989(2) | 1.86558(2) | 1.79533(2) |
| Ge2-O7 (Å) | 1.67937(2) | 1.68963(3) | 1.81800(5) | 1.72130(2) | 1.61730(2) | 1.47662(2) |
| Ge2-O4-Ge1 (°) | 118.430(1) | 118.990(1) | 116.100(2) | 114.975(1) | 113.143(1) | 111.155(1) |
| Ge1-O3-Ge1 (°) | 137.562(1) | 135.969(1) | 131.959(2) | 144.589(1) | 133.630(1) | 137.072(1) |
| Ba-O1(Å) | 2.76820(3) | 2.75588(5) | 2.74590(8) | 2.74857(4) | 2.8039(4) | 2.89752(4) |
| Ba-O1 (Å) | 3.20360(3) | 3.16168(5) | 3.18610(9) | 3.11353(3) | 3.0377(5) | 3.25055(4) |
| Ba-O2 (Å) | 2.76802(3) | 2.83622(5) | 2.86321(8) | 2.41572(3) | 2.8939(4) | 2.97082(4) |
| Ba-O2 (Å) | 2.95525(3) | 2.81563(4) | 2.70616(6) | 3.31386(3) | 2.7050(4) | 2.51634(3) |
| Ba-O3 (Å) | 2.85206(3) | 2.83676(4) | 2.79753(6) | 2.92805(3) | 2.82424(3) | 2.84378(3) |
| Ba-O4 (Å) | 3.11084(2) | 3.09398(4) | 3.06646(6) | 3.00537(3) | 2.94699(3) | 2.96462(3) |
| Ba-O5 (Å) | 2.87023(4) | 2.83514(5) | 2.80777(8) | 3.04164(5) | 3.2981(4) | 2.96208(5) |
| Ba-O6 (Å) | 3.02353(3) | 2.97544(5) | 3.18379(8) | 3.32304(4) | 2.6959(4) | 3.06210(4) |
| Ba-O6 (Å) | 2.87635(3) | 2.92350(5) | 2.73258(7) | 2.59942(3) | 2.9756(5) | 2.85667(4) |
| Ba-O7 (Å) | 3.00306(3) | 3.02879(5) | 3.04686(8) | 2.82659(4) | 2.9636(4) | 2.88366(4) |

## 3.2. Magnetic Properties of Ba$_2$RE$_2$Ge$_4$O$_{13}$ Polycrystals.

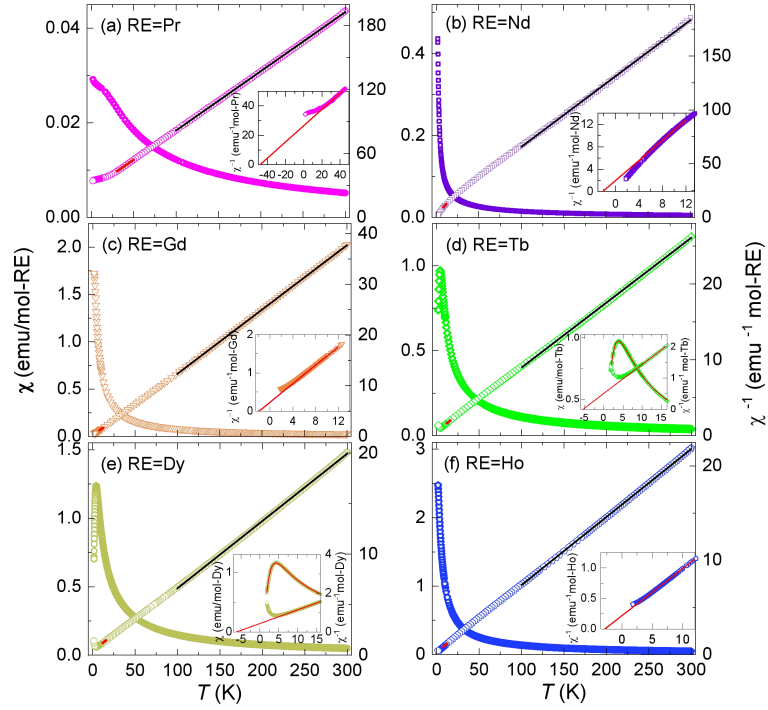

**Figure 3**. (a)-(f) Temperature dependence of *dc* magnetic susceptibility $\chi(T)$ and inverse susceptibility $1/\chi(T)$ measured at $\mu_0H$ = 0.1 T for Ba$_2$RE$_2$Ge$_4$O$_{13}$ (RE = Pr, Nd, Gd-Ho), respectively. The black and red lines show the CW fitting at high-temperature and low-temperature regimes. The inset shows the low temperature $1/\chi(T)$ curves, the red lines in (d, e) show the fits of spin-dimer model to the $\chi$(T) data of Ba$_2$RE$_2$Ge$_4$O$_{13}$ (RE = Tb, Dy).

Temperature dependence of *dc* magnetic susceptibilities $\chi(T)$ for the series of Ba$_2$RE$_2$Ge$_4$O$_{13}$ (RE = Pr, Nd, Gd-Ho) samples were measured from 1.8 to 300 K with an applied field $\mu_0H$ = 0.1 T, as shown in Figure 3. Among the Ba$_2$RE$_2$Ge$_4$O$_{13}$ series, the $\chi(T)$ curves of two compounds with RE = Tb, Dy show the broad maximum at ~ 3.7 K (RE = Tb) and ~ 3.9 K(RE = Dy), indicating the development of short-range magnetic correlations usually observed in the frustrated spin systems or the gapped spin-singlet system (see Figure 3d,e). For the other compounds, no magnetic anomaly is detected in the $\chi(T)$ curves at temperatures above 1.8 K. The inverse susceptibility $1/\chi(T)$ results were fitted by the Curie-Weiss (CW) law, $\chi = C/(T − \theta_{CW})$, where $\theta_{CW}$ is the Curie-Weiss temperature and $C$ is the Curie Constant. The effective magnetic moments $\mu_{eff}$ were calculated by the following relationship: $\mu_{eff} = (3k_BC/N_A)^{1/2}$, where $k_B$ is the Boltzmann constant and $N_A$ is Avogadro's constant. The $1/\chi(T)$ curves are fitted at both high temperature and low temperature regimes, the resultant $\theta_{CW}$ and $\mu_{eff}$ from the CW fitting are listed in Table 2, where the effective moments ($\mu_{fi}$) of free RE$^{3+}$ ions are also presented for comparisons. At high temperatures (*T* >

100 K), $1/\chi(T)$ curves nearly follow the linear temperature dependence, and the obtained effective moment $\mu_{eff}$ is close to the theoretical free-ion value $\mu_{fi}$ calculated by $g_J[J(J+1)]^{1/2}\mu_B$, where $g_J$ is Landé factor, $J$ is the total angular moment. As $T$ decreases below 50 K, the slope of $1/\chi(T)$ curves is changed and which is more clearly observed for RE = Pr and Nd samples due to the thermal population of electrons at different crystal electric field (CEF) levels. Thus, the fitted $\theta_{CW}$ and $\mu_{eff}$ at low temperature regimes ($T < 16$ K) can better reflect the magnetic parameters of ground states, where more population of electrons occupy on the lowest-lying CEF levels. This can explain why the low-T fitted $\mu_{eff}$ are smaller than the high-T ones. For RE = Pr, we can notice that the $\chi(T)$ curves exhibit a broad shoulder-shaped behavior below 20 K and without distinct increment of magnetization as decreased temperatures. This behavior can be related to the formation of two low-lying singlets with sizable energy separation for the $Pr^{3+}$ ($4f^2$, $^3H_4$) non-Kramers ions, as the report in $Pr_3M_2Nb_3O_{14}$ (with M = Mg or Zn) and $PrMAl_{11}O_{19}$ (M = Mg or Zn) compounds.[48,49] For the other compounds, the low-T negative values of $\theta_{CW}$ can de related to the dominant AFM exchange interactions between local $RE^{3+}$ moments. As an example, $\theta_{CW}$ = -1.89 K in $Ba_2Gd_2Ge_4O_{13}$ suggests the dominant AFM exchange interactions between $Gd^{3+}$ moments since the $Gd^{3+}$ ions are half-filled 4f shell ($4f^7$, $S = 7/2$, $L = 0$) with quasi-isotropic magnetic interactions.

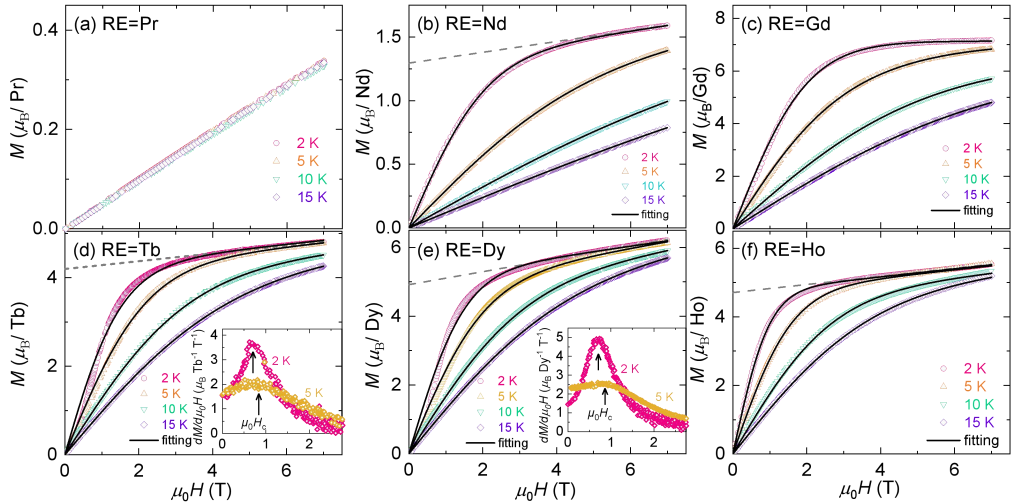

**Figure 4.** (a)-(f) The field dependent magnetization $M(\mu_0H)$ curves at selected temperatures for $Ba_2RE_2Ge_4O_{13}$ (RE = Pr, Nd, Gd-Ho). The dashed line indicates the Van Vleck paramagnetic contribution extracted from the linear-field-dependent magnetization data at high field regimes. The inset in (d, e) show the derivative magnetization $dM/d\mu_0H$ curves of $Ba_2RE_2Ge_4O_{13}$ (RE = Tb, Dy), $\mu_0H_C$ is defined by the peak position of derivative magnetization $dM/d\mu_0H$.

The isothermal field-dependent magnetization $M(\mu_0H)$ curves at selected temperatures for $Ba_2RE_2Ge_4O_{13}$ (RE = Pr, Nd, Gd-Ho) are shown in Figure 4a-f, respectively. For RE = Pr, the $M(\mu_0H)$ curves display the linear field dependence behaviors with maximized magnetization $M_S$~0.34 $\mu_B$/Pr$^{3+}$ at 7 T, this value is far smaller than the saturated magnetization for the system with $J_{eff}$ = 1/2 effective moment but in line with the formation of well-separated singlet state of non-Kramers Pr$^{3+}$ ($4f^2$, $^3H_4$) ions. The saturated magnetization of Ba$_2$Gd$_2$Ge$_4$O$_{13}$ at 2 K reaches ~7.13 $\mu_B$, this is close to $M_S$ = 7$\mu_B$/Gd$^{3+}$ for the half-filled Gd$^{3+}$ ($4f^7$) ions. For the others, the $M(\mu_0H)$ curves at 2 K reach the saturation up to field $\mu_0H$ = 7 T, the linear increase of magnetization in high field regimes ($B \geq 4$ T) arises from the Van-Vleck (VV) paramagnetic (PM) contribution. After subtracting the VV paramagnetic contribution, the obtained saturated magnetizations at 2 K are also presented in Table 2, which are smaller than the values of free RE$^{3+}$ ions. For RE = Tb and Dy, the $M(\mu_0H)$ curves exhibit the field-induced spin flop behaviors at temperatures below $T_p$ (see the inset of Figure 4d,e), where the critical field is defined by the peak position of derivative magnetization d$M$/d$\mu_0H$ results.

To understand the magnetic behaviors of Ba$_2$RE$_2$Ge$_4$O$_{13}$, firstly we should point out that RE$^{3+}$ ions can be divided into Kramers ions (such as Nd$^{3+}$, Dy$^{3+}$) and non-Kramers ions (like Pr$^{3+}$, Tb$^{3+}$, and Ho$^{3+}$) in terms of the number of 4$f$ electrons being odd or even,[50,51] which can affect the split CEF levels and low-$T$ ground states. For the compounds containing Kramers RE$^{3+}$ ions, the low-$T$ magnetism can be well described by an effective $J_{eff}$ = 1/2 moment protected by time-reversal symmetry. While, the situation for non-Kramers RE$^{3+}$ ions is more complex, its ground state is affected by the local coordination environment of the RE$^{3+}$ ions, two nonmagnetic CEF singlets are usually formed. In the low crystal field symmetry, these two low-lying CEF singlets have a small energy gap, thus the ground state can be described by a quasi-doublet ground state with a pseudospin-1/2 moment. Here, for Ba$_2$Pr$_2$Ge$_4$O$_{13}$ compound, the local $C_1$ symmetry environment in PrO$_7$ polyhedra leads to the well separated two singlet states, which can explain the small magnetization values at 2 K and 7 T. In case of RE= Tb$^{3+}$ ($4f^6$, J = 6) and Ho$^{3+}$($4f^8$, J = 8), the saturation magnetization at 7 T reaches the value of half saturated magnetization $M_{Sat}$ = $g_JJ\mu_B$/2, here $g_J$ is the Lande's factor and $\mu_B$ is the Bohr magneton. This indicates the gap between the two singlet states is small, thus the low-T magnetic behaviors can be described by the quasi-doublet ground states. Except for RE = Pr, the isothermal $M(\mu_0H)$ curves for Ba$_2$RE$_2$Ge$_4$O$_{13}$ can be fitted by the Brillouin function $B_J(x) = \frac{2J_{eff}+1}{2J_{eff}} \coth \frac{2J_{eff}+1}{2J_{eff}} x - \frac{1}{2J_{eff}} \coth \frac{x}{2J_{eff}}$ where $x = g\mu_BJ_{eff}\mu_0H/k_BT$. For Ba$_2$Gd$_2$Ge$_4$O$_{13}$ ($J_{eff}$ = 7/2), the fits to the $M(\mu_0H)$ curves at 2 K yield $g$-factor is ~2.08 (3). For the others, the $g$-factor as adjusting parameter can be obtained from the fitting to the experimental data by fixing $J_{eff}$ = 1/2. As examples, the $g$-factors are determined to ~2.41 (3) (RE = Nd) and ~9.71 (2) (RE = Dy) at 2 K.

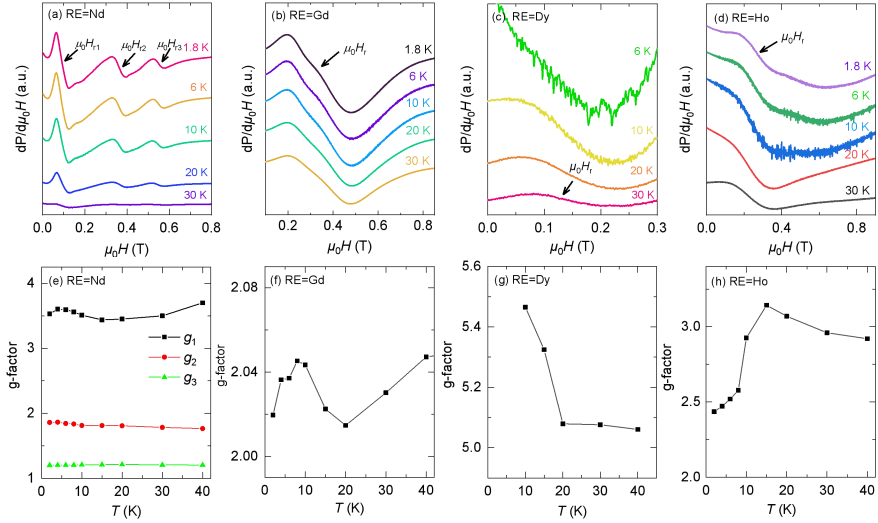

**Figure 5**. (a-d) The X-band electron spin resonance (ESR) spectra at selected temperatures, (e-h) the obtained $g$-factors for $Ba_2RE_2Ge_4O_{13}$ (RE = Nd, Gd, Tb, Dy, Ho).

The X-band electron spin resonance (ESR) measurements for $Ba_2RE_2Ge_4O_{13}$ (RE = Nd, Gd, Dy,Ho) were carried out at different temperatures. The derivative ESR spectra ($dP/d\mu_0H$) ($P$ is the integral ESR intensity) are presented in Figure 5a-d. From that, the $g$-factors are calculated as $g = h\nu/\mu_B\mu_0H_r$, where $h$ is the Planck constant, $\nu$ = 9.4 GHz is the microwave frequency, $\mu_0H_r$ is the resonance absorption field and $\mu_B$ is the Bohr magneton. The obtained temperature-dependent Lande $g$-factors for are shown in Figure 5e-h. For RE = Nd, three sets of resonance lines can be identified possibly due to the low symmetry of distorted $NdO_7$ coordinations, which give the $g$-factors of $g_1$ = 3.53, $g_2$ = 1.86 and $g_3$ = 1.20 at 2 K. Then, using the averaged $g$-factor ($g_{ave}$) calculated by $g_{ave} = \sqrt{(g_1^2 + g_2^2 + g_3^2)/3}$, $M_S$ is calculated to be $M_S$ = $g_{ave}J_{eff}\mu_B$ ~ 1.21$\mu_B$, which matches well with the experimental value $M_S$ = 1.28$\mu_B$. For RE = Gd, the estimated $g$-factor ~2.02 at 2 K yield $M_{Sat}$ = 7.04 $\mu_B$, which is close to the experimental value $M_S$ = 7.13$\mu_B/Gd^{3+}$. For RE = $Dy^{3+}$ and $Ho^{3+}$, the obtained $g$-factors $g$ = 5.0-5.5 and $g$ = 2.4-3.2 give the $M_S$ ~ 2.5-2.75 $\mu_B/Dy^{3+}$ and $M_S$ ~ 1.2-1.6$\mu_B/Ho^{3+}$ much smaller than the experimental values, thus the other influencing factors including the contribution of orbital moments and local CEF environments of $Dy^{3+}/Ho^{3+}$ ions should be considered for this deviation.

Below, we evaluate the intra-ladder, inter-ladder dipolar interactions and the superexchange interactions of $Ba_2RE_2Ge_4O_{13}$. The dipolar interaction ($D$) is evaluated by $D = \mu_0\mu_{eff}^2/4\pi(r_{nn}^3)$,[47,51] where $r_{nn}$ refers to intra-ladder distances ($d_{rung}$, $d_{leg}$) and NN inter-ladder distances ($d_{inter,a}$, $d'_{inter,a}$) as denoted in Figure 1d. Here, the estimated $D$ is assumed to be isotropic and only include the NN exchange interactions, thus it only provides a rough estimate of D in the energy scale and cannot reflect the single-ion anisotropy of $RE^{3+}$ ions. Using the low-T fitted magnetic parameters $\theta_{CW}$ and $\mu_{eff}$, the calculated dipole interactions are listed in Table 2. From that, the dipolar interactions along the rung ($D_{rung}$) are ~3 times larger than the ones along the leg ($D_{leg}$). Compared to the typical Dy-based magnet $Dy_2Ti_2O_7$ with $D$ ~ 1.41 K,[47,52] the $D_{rung}$ of $Ba_2Dy_2Ge_4O_{13}$ has a comparable value ~1.46 K. The strength of superexchange interactions between local $RE^{3+}$ moments are estimated by the mean field approximation using $J_{nn} = 3k_B\theta_{CW}/zS_{eff}(S_{eff}+1)$,[53] where $S_{eff}$ denotes the total spin quantum number, and z is the number of NN spins (z = 1 for the rung interactions). Here, both quantum number $J = |L \pm S|$ and effective spin number $S_{eff} = 1/2$ of $RE^{3+}$ ions are employed to estimate the rung interaction $J_{rung}$, the results are shown in Table 2. For compounds containing the Kramers ions, it is more suitable to take $S_{eff} = 1/2$ and that is applicable to the $Ba_2RE_2Ge_4O_{13}$ (RE = Nd, Dy), where the electrons mainly occupy on the low-lying Kramers doublet states at low temperatures. It is also to be noted that the mean-field approximation of $J_{nn}$ does not take account for the local single-ion anisotropy and other exchange interactions of $RE^{3+}$ ions, and the value of $\theta_{CW}$ depends on the fitted temperature ranges. More accurate determinations of $D$ and $J_{nn}$ require future inelastic neutron spectroscopy (INS) measurements on the $Ba_2RE_2Ge_4O_{13}$ compounds, which can determine the spin anisotropy and CEF information. From the structure viewpoint, due to the much larger super-exchange distances through the RE-O-Ba/Ge-O-RE pathways, both $J_{leg}$ and inter-ladder ($J_{inter}$) exchange interactions should be much smaller than $J_{rung}$ via the RE-O-RE routes.

**Table 2.** The Curie-Weiss temperatures ($\theta_{CW}$) and effective magnetic moments ($\mu_{eff}$) determined from the Curie-Weiss fitting at low and high temperatures to magnetic susceptibility $\chi(T)$ of $Ba_2RE_2Ge_4O_{13}$ (RE = Pr, Nd, Gd-Ho) compounds, the effective moment ($\mu_{fi}$) for free ions are calculated by $g[J(J + 1)]^{1/2}$, the obtained saturated magnetizations at 2 K ($M_{S,2K}$), the calculated superexchange interaction ($J_{nn}$), the estimated rung ($D_{rung}$), leg ($D_{leg}$) and inter-ladder ($D_{inter}$) exchange couplings.

| RE | High T fit | $\theta_{CW}$(K) | $\mu_{eff}(\mu_B)$ | Low T fit | $\theta_{CW}$(K) | $\mu_{eff}(\mu_B)$ | $\mu_{fi}(\mu_B)$ | $M_{S,2K}(\mu_B)$ | $J_{rung}$ $J_{nn}$ using $J$ (K) | $J_{nn}$ using $S_{eff}=1/2$ (K) | $D_{rung}$(K) | $D_{leg}$(K) | $D_{inter}$(K) |
|---|---|---|---|---|---|---|---|---|---|---|---|---|---|
| Pr | 100-300 K | -46.7 | 3.78 | 30-50 K | -48.21 | 3.81 | 3.58 | 0.34 | N/A | N/A | 0.165 | 0.056 | 0.042 |
| Nd | 100-300 K | -10.67 | 3.68 | 8-12 K | -2.19 | 2.98 | 3.62 | 1.30 | -0.13 | -4.38 | 0.103 | 0.035 | 0.040 |
| Gd | 100-300 K | 2.56 | 7.95 | 8-12 K | -1.89 | 8.10 | 7.94 | 7.13 | -0.18 | N/A | 0.754 | 0.263 | 0.195 |
| Tb | 100-300 | -1.95 | 9.55 | 12-16 | -5.48 | 9.29 | 9.72 | 4.12 | -0.196 | -10.96 | 1.05 | 0.348 | 0.276 |

| | | | | | | | | | | | | |
|---|---|---|---|---|---|---|---|---|---|---|---|---|
| | K | | | K | | | | | | | | |
| Dy | 100-300 K | -0.86 | 10.62 | 12-16 K | -6.40 | 10.68 | 10.63 | 4.91 | -0.151 | -12.8 | 1.462 | 0.462 | 0.338 |
| Ho | 100-300 K | -2.03 | 10.53 | 8-12 K | -2.82 | 10.18 | 10.60 | 4.71 | -0.059 | -5.64 | 1.302 | 0.422 | 0.348 |

To get more information on magnetic behaviors of Ba$_2$Dy$_2$Ge$_4$O$_{13}$, the *dc* magnetic susceptibilities under different magnetic fields were measured, as shown in Figure. 6a. A broad maximum is observed in the $\chi(T)$ curves at low fields (≤ 0.5 T) and it is gradually suppressed for field above 0.7 T. Also, the real part of *ac* susceptibility ($\chi'_{ac}$) data at various frequencies from ~ 9 Hz to 997 Hz in zero dc magnetic field is shown in Figure 6b. At low frequency $f$ = 9 Hz, only a single broad peak maximized at $T_{p1}$ ~ 4.2 K is observed corresponding to the broad peak observed in the low-field static susceptibility. As the frequency increases, another high temperature broad peak at $T_{p2}$ is observed, and $T_{p2}$ shows a frequency dependence. As shown in Figure S3, the frequency dependence of $T_{p2}$ follows the Arrhenius law, [54,55] suggesting the presence of slow magnetic relaxation behavior as the observation in the Dy-based single molecule magnet and spin ice systems. [55-57] By contrast, the position of $T_{p1}$ doesn't change with the frequency. From the above dc- and ac-susceptibility, no sharp peak is detected indicating the absence of long-range magnetic order at temperature above 1.8 K.

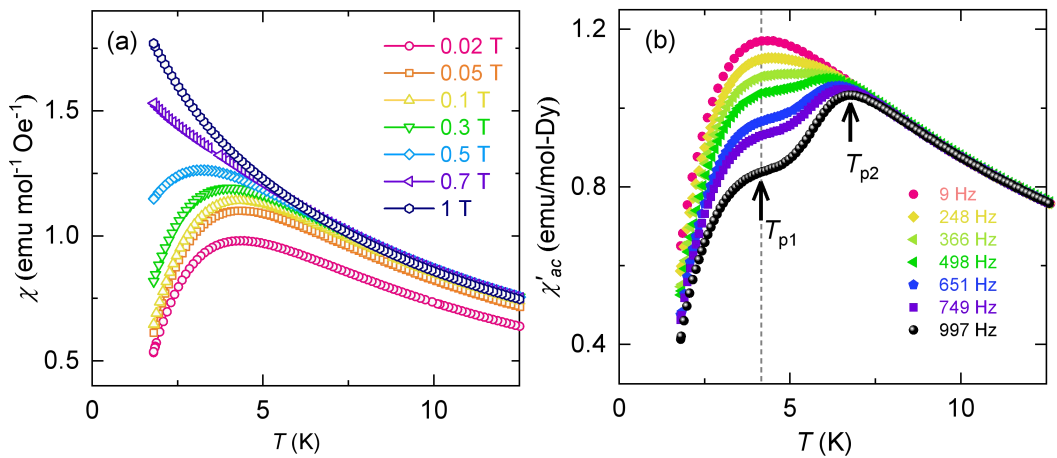

**Figure 6** (a) Low-temperature *dc* magnetic susceptibility of Ba$_2$Dy$_2$Ge$_4$O$_{13}$ at different magnetic fields, (b) the real part of *ac* susceptibility of Ba$_2$Dy$_2$Ge$_4$O$_{13}$ at different frequencies, the $T_{p1}$ and $T_{p2}$ are denoted by arrows, and dashed line indicates the peak position of $T_{p1}$.

In the strong-rung regime, the two-leg spin ladder can show the spin-dimer behavior due to the two coupled spins along the rung in spin-1/2 ladder.[46,58] Then, the χ(T) data of Ba$_2$RE$_2$Ge$_4$O$_{13}$ (RE = Tb, Dy) is fitted by a coupled spin dimer model:[59,60]

$$\chi(T) = \chi_0 + \frac{N_A g^2 \mu_B^2}{k_B T[3 + \exp(J/k_B T) + zJ'/k_B T]},$$ where the first term represents the Van Vleck PM contribution, the second term is the contribution from interacting spin dimers. This model can well describe the strong-rung spin ladder systems, in which the $N_A$, $\mu_B$, $k_B$, and $g_J$ denote Avogadro's number, Bohr magneton, Boltzmann constant, and Landé factor, respectively. Here, $J/k_B$ is the rung interactions, $J'/k_B$ represents the exchange interactions of both $J_{leg}$ and $J_{inter}$ and $z$ is the coordinated number of NN dimers. As shown in Figure 3d,e and Figure 7a, this model can give a good fit to the χ(T) curves with $J/k_B$ = 6.99(1) K ($J/k_B$ = 5.77(2) K) and $J'/k_B$ = 0.61(3) K ($J'/k_B$ = 0.67(3) K) for RE = Dy (Tb) [ see the details in Figure S4]. Moreover, using $J/k_B$ = 6.99 K and $g$ = 9.28(3) of Ba$_2$Dy$_2$Ge$_4$O$_{13}$, a critical field ($\mu_0 H_C$) for closing the spin gap ($\Delta_0 = J/k_B$) is to $\mu_0 H_C = \Delta_0/g\mu_B$ = 0.62 T, this is nearly in accordance with the $M(\mu_0 H)$ results [see Figure 7b]. Also, the ratios of $J/J'$ = 11.47 (RE = Dy) and $J/J'$ = 8.61 (RE = Tb) support the Ba$_2$RE$_2$Ge$_4$O$_{13}$ can be considered as a strong-rung spin ladder system.

### 3.3. The Ultra-Low Temperature Magnetization and Specific Heat of Ba$_2$Dy$_2$Ge$_4$O$_{13}$

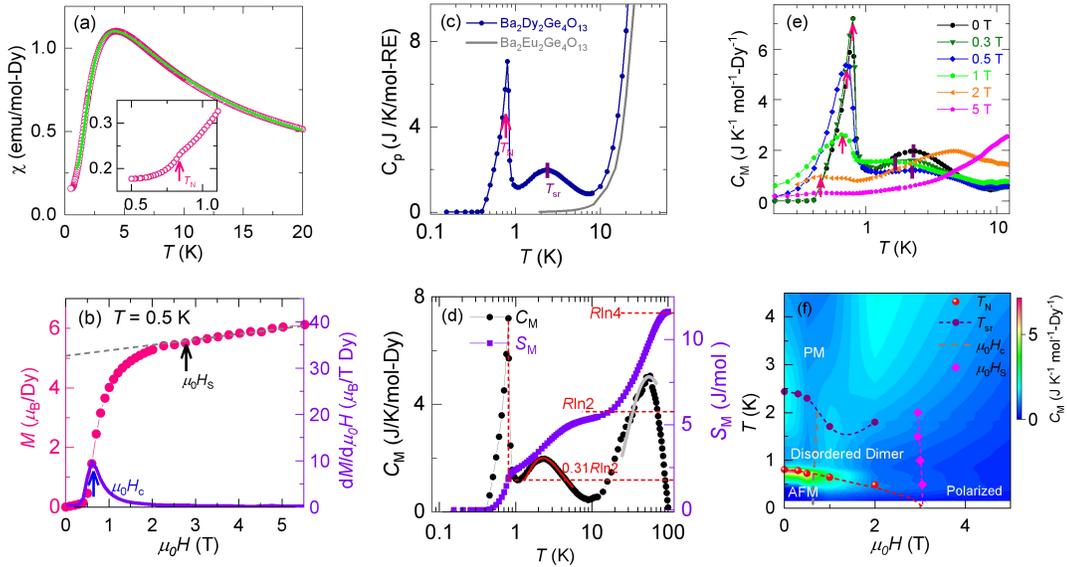

**Figure 7.** (a) Temperature dependence of *dc* magnetic susceptibility χ(T) curves of Ba$_2$Dy$_2$Ge$_4$O$_{13}$, the green lines show the fits by coupled spin dimer model, the inset shows the low-temperature part of χ(T) curves. (b) The field dependent magnetization $M(\mu_0 H)$ and derivative magnetization $dM/d\mu_0 H$ curves at 0.5 K for Ba$_2$Dy$_2$Ge$_4$O$_{13}$. (c) The zero-field specific heat $C_p(T)$ curves of Ba$_2$Dy$_2$Ge$_4$O$_{13}$, the gray lines show the $C_p(T)$ data of nonmagnetic Ba$_2$Eu$_2$Ge$_4$O$_{13}$. (d) The zero field magnetic specific heat $C_M(T)$ and magnetic entropy $S_M(T)$ results of Ba$_2$Dy$_2$Ge$_4$O$_{13}$. (e) Temperature dependence of $C_M(T)$ curves under different fields. (f) The magnetic phase diagram of Ba$_2$Dy$_2$Ge$_4$O$_{13}$ showing the PM phase, the disordered dimer state and AFM

phase. The AFM phase consists of two regions, the low-field AFM phase and canted AFM phase at intermediate field regimes separated by the critical field ($\mu_0H_C$). At low temperature and high field ($\mu_0H > \mu_0H_S$) regions, it enters into the fully polarized magnetic phase.

Among the $Ba_2RE_2Ge_4O_{13}$ family members, $Ba_2Dy_2Ge_4O_{13}$ exhibits the strong magnetic interactions with large values of $\theta_{CW}$ = -6.4 K and $T_{p1}$ ~ 3.9 K indicated from the $\chi(T)$ data, thus it provides a good system to investigate the strong-rung spin ladder physics with $J_{eff}$ = 1/2 Kramers doublet state. To reveal its magnetic ground state, zero-field specific heat $C_p(T)$ of $Ba_2Dy_2Ge_4O_{13}$ was measured with temperature down to 0.1 K. As presented in Figure 7c, the typical feature is the coexistence of a broad peak centered at $T_p$ ~ 2.4 K and a large λ-type anomaly occurred at $T_N$ ~ 0.81 K, the former corresponds to the formation of disordered dimer-singlet state and the latter indicates the transition to a long-range magnetic ordered phase. Here, the former is related to the strong rung interaction, and the long-range AFM order can be ascribed to the establishment of intra-leg interaction as well as inter-ladder interactions at reduced temperatures, similar to the quasi-1D chain or spin-dimer systems due to the enhanced inter-chain or inter-dimer exchange interactions at decreased temperatures as observed in 1D AFM spin chain $K_2PbCu(NO_2)_6$[61] and spin dimer $LiCu_2O_2$ compounds.[62] Using the specific heat of isostructural nonmagnetic $Ba_2Eu_2Ge_4O_{13}$ as the lattice contribution $C_{Latt}(T)$, magnetic specific heat $C_M(T)$ is singled out from the total specific heat data, as plotted in Figure 7d. Using the isolated spin dimer model, the energy gap Δ~ 6.86 K is obtained from the fitting to the zero-field $C_M(T)$ data as shown in Figure 7d and Figure S4, which is close to the value from the susceptibility data. By integrating $C_M/T$ over temperature, magnetic entropy $S_M(T)$ data is also obtained. Up to ~ 15 K, the $S_M(T)$ approaches to a well-defined plateau with a value of ~0.92Rln2 as expected for the effective $J_{eff}$ = 1/2 states, where R is the gas constant. While, only ~ 31% of magnetic entropy is released below $T_N$, this means that ~ 60% of magnetic entropy is released at temperatures well above $T_N$ due to the disordered dimer state. Additionally, besides the two peaks at $T_N$ and $T_p$, a high-T broad hump maximized at ~ 60 K is visible, which is in associated with the Schottky anomaly with CEF energy splitting of the J = 15/2 multiplet of $Dy^{3+}$ ($4f^9$, $^6H_{15/2}$) ions. The $S_M(T)$ reaches ~Rln4 at T = 100 K indicative of the first excited doublet states. Using the two-level Schottky formula [see the grey lines in Figure 7d],[63] the energy gap of the first excited CEF level separated from the doublet ground state is determined to be Δ ~ 168 K, this large energy gap ensures the well separated Kramers doublet ground states from the 1$^{st}$ excited state. Thus, $Ba_2Dy_2Ge_4O_{13}$ can be well considered as an effective $J_{eff}$ = 1/2 moment system in the low temperature and low field regimes.

Under applied magnetic field, both $T_N$ and $T_p$ shift to the low temperatures, where the peak positions are denoted by grey arrows and red vertical lines in Figure 7e. For $\mu_0H$ > 2 T, another high temperature Schottky anomaly emerges and which smooths down the broad peak of short range spin correlations. At field of 5 T, $T_N$ is suppressed down below 0.2 K from $T_N$~0.81 K at zero field. Using the ordering temperature ($T_p$, $T_N$) and critical fields ($\mu_0H_c$, $\mu_0H_S$) obtained from the $C_M$(T) and $M(\mu_0H)$ data, the $\mu_0H$-T magnetic phase diagram is constructed as shown in Figure 7f. In the phase diagram, $Ba_2Dy_2Ge_4O_{13}$ experiences a crossover from a disordered dimer singlet state to an AFM state at zero field. The AFM phase resides at low temperature ($T \leq T_N$) and field below $\mu_0H_S$, and this phase can be further separated into the low-field AFM phase by ($\mu_0H \leq \mu_0H_C$) and canted AFM phase at intermediate fields ($\mu_0H_c \leq \mu_0H \leq \mu_0H_S$). As $\mu_0H > \mu_0H_S$, it enters into the full polarized magnetic state. Additionally, the disordered dimer singlet phase is located at intermediate temperature ($T_N \leq T \leq T_p$) and field of $\mu_0H \leq \mu_0H_S$ regimes.

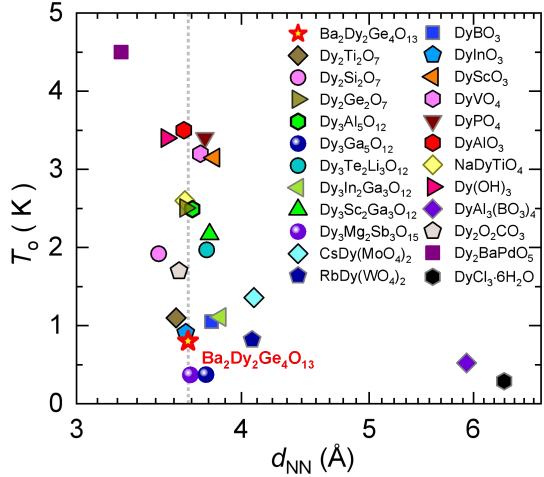

**Figure 8** The ordering temperature ($T_o$) versus the distance of nearest-neighbor $Dy^{3+}$ ions ($d_{NN}$) of Dy-based oxides, the dashed line denotes the position of $d_{NN}$ ~ 3.64 Å of $Ba_2Dy_2Ge_4O_{13}$.

Lastly, to illustrate the effect of low dimensionality of spin ladder on magnetic ordering temperature ($T_o$) in $Ba_2Dy_2Ge_4O_{13}$, we compare its $T_o$ with the ones of other insulating Dy-based magnets reported in the literatures.[64-84] For the insulating Dy-based compounds, the super-exchange and dipolar interactions between $Dy^{3+}$ ions strongly depend on the NN distance ($d_{NN}$) between $Dy^{3+}$ ions, thus we plot the $T_o$ versus $d_{NN}$ for the Dy-based materials as shown in Figure 8. Among the Dy-based compounds with reported $T_o$, the low $T_o$~0.52 K of $DyAl_3(BO_3)_4$ with $d_{NN}$ = 5.928 Å[82] and $T_o$~0.289 K of $DyCl_3·6H_2O$ with $d_{NN}$ = 6.33 Å[72] indicate a general trend that larger $d_{NN}$ will lead to the reduced $T_o$. We also can find that, compared to the Dy-based magnets with comparable $d_{NN}$ ~3.65 Å, $Ba_2Dy_2Ge_4O_{13}$ exhibits a much lower $T_o$ ~ 0.81 K. This result reveals that the low dimensionality of spin ladder motif indeed plays a significant role on reducing $T_o$. Another two compounds with much reduced $T_o$ are $Dy_3Ga_5O_{12}$ and $Dy_3Mg_2Sb_3O_{14}$,[65,73,79] both compounds belong to the geometrical frustrated magnets with Kagome lattice. And Indeed, recent studies have revealed that the reduced $T_o$ and the quantum spin liquid state namely without long-range magnetic order can be realized by the enhanced spin frustration effect, which can explain the case of $Dy_3Ga_5O_{12}$ and $Dy_3Mg_2Sb_3O_{15}$. By comparison, here the low $T_o$ and absence of geometric frustration highlight the effect of low dimensionality on suppressing $T_o$, as another effective scenario to reduce the $T_o$ of quantum magnets. In all, $Ba_2Dy_2Ge_4O_{13}$ provides a rare spin ladder system with effective $J_{eff}$ = 1/2 magnetic moments, and which has the coexistence of disordered dimer and long-range AFM states.

## 4.CONCLUSION

In summary, we have reported a new family of two-leg spin ladder compounds $Ba_2RE_2Ge_4O_{13}$ (RE = Pr, Nd, Gd-Ho), where the magnetic RE ions are connected via the RE-O-RE and RE-O-Ge-O-RE exchange pathways with $d_{rung}$ = 3.64-3.79 Å and $d_{leg}$ = 5.34-5.43 Å along the rungs and legs of ladder, respectively. Bulk magnetization characterizations on all synthesized compounds reveal the dominant AFM interactions and absence of long-range magnetic order at temperatures down to 1.8 K, and the $\chi(T)$ curves exhibit the broad maximum at ~ 3.7 K (RE = Tb) and ~ 3.9 K (RE = Dy) indicative of the formation of spin gapped dimer singlet state due to the strong rung interactions. Moreover, as one family member with $J_{eff}$ = 1/2 Kramers doublet states, $Ba_2Dy_2Ge_4O_{13}$ exhibits a phase transition from the spin-dimerized singlet state to the long-range AFM order at $T_N$ = 0.81 K associated with the enhanced intra-leg or inter-ladder exchange couplings at reduced temperatures. Our experimental results establish the family of $Ba_2RE_2Ge_4O_{13}$ as an attractive platform for exploring the exotic magnetic states in strong-rung spin ladder materials consisting of RE ions.

**ACKNOWLEDGMENTS**

The work is supported by the National Natural Science Foundation of China (Grant No. 11874158), and the National Key Research and Development Program (Grant No. 2023YFA1406500 & 2021YFC2202300). Part of this work was supported by the Guangdong Basic and Applied Basic Research Foundation (Grant No.2022B1515120020) and the Fundamental Research Funds of Guangdong Province (Grant No. 2022A1515010658). A portion of measurements was carried out at the synergetic extreme condition user facility (SECUF) and the Steady High Magnetic Field Facilities. We would like to thank Huifen Ren for her assistance on the magnetic measurement and thank the staff of the analysis center of

**References**

(1) Vasiliev, A.; Volkova, O.; Zvereva, E.; Markina, M. Milestones of low-D quantum magnetism, *npj quantum Materials*. **2018**, 3,18.

(2) Dagotto, E.; Rice, T.M. Surprises on the way from one- to two-dimensional quantum Magnets: The Ladder Materials. *Science*, **1996**, 271, 618-623.

(3) Ising, E. Report on the theory of ferromagnetism. *Z. Phys*. **1925**, 31, 253-258.

(4) Bethe, H. On the theory of metals: Ⅰ. Eigenvalues and eigenfunctions for the linear atomic chain. *Z. Phys*. **1931**, 71, 205-226.

(5) Mermin, N. D.; Wagner, H. Absence of ferromagnetism or antiferromagnetism in one-or two-dimensional isotropoic Heisenberg models. *Phys. Rev. Lett.* **1966**,17,1133.

(6) Manousakis, E. The spin-1/2 Heisenberg antiferromagnet on a square lattice and its application to the cuprous oxides. *Rev. Mod. Phys.* **1991**, 63, 1.

(7) Förster, T.; Garcia, F. A.; Gruner, T.; Kaul, E. E.; Schmidt, B.; Geibel, C.; Sichelschmidt, J. Spin Fluctuations with Two-Dimensional XY Behavior in a Frustrated S = 1/2 Square-Lattice Ferromagnet, *Phys. Rev. B.* **2013**, 87, 180401.

(8) Hong, T.; Kim, Y.H.; Hotta, C.; Takano, Y.; Tremelling, G.; Turnbull, M. M.; Landee, C. P.; Kang, H.-J.; Christensen, N. B.; Lefmann, K.; Schmidt, K. P.; Uhrig, G. S.; Broholm, C. Field-induced Tomonaga-Luttinger Liquid phase of a two-leg spin-1/2 Ladder with strong leg interactions, *Phys. Rev. Lett.* **2010**, 105, 137207.

(9) Anderson, P. W. The resonating valence bond state in $La_2CuO_4$ and superconductivity. *Science*, **1987**, 235, 1196.

(10) Johnston, D.C.; Johnson, J. W.; Goshorn, D. P.; Jacobson, A. J. Magnetic susceptibility of $(VO)_2P_2O_7$: a one-dimensional spin-1/2 Heisenberg antiferromagnet with a ladder spin configuration and a singlet ground state. *Phys. Rev. B.* **2013**, 87, 064404.

(11) Thurber, K. R.; Imai, T.; Saitoh, T.; Azuma, M.; Takano, M.; Chou, F. C. $^{63}$Cu NQR Evidence of Dimensional Crossover to Anisotropic 2D Regime in S = 1/2 Three-Leg Ladder $Sr_2Cu_3O_5$. *Phys. Rev. Lett.* **2000**, *84* (3), 558–561.

(12) A. Ghosh and I. Bose, Antiferromagnetic Spin Ladders with Odd and Even Numbers of Chains, *Phys. Rev. B.* **55**, 3613 (1997).

(13) Schwenk, H.; König, D.; Sieling, M.; Schmidt, S.; Palme, W.; Lüthi, B.; Zvyagin, S.; Eccleston, R. S.; Azuma, M. Takano. M. Magnetic resonances in spin ladder systems $(VO)_2P_2O_7$, $SrCu_2O_3$ and $Sr_2Cu_3O_5$, *Phys B: Condensed Matter* **1997**, 237-238, 115-116.

(14) Watson, B. C.; Kotov, V. N.; Meisel, M. W.; Hall, D. W.; Granroth, G. E.; Montfrooij, W. T.; Nagler, S. E.; Jensen, D. A.; Backov, R.; Petruska, M. A.; Fanucci, G. E.; Talham, D. R. Magnetic spin Ladder $(C_5H_{12}N)_2CuBr_4$: High-field magnetization and scaling near quantum criticality. *Phys. Rev. Lett.* **2001**, 86, 5168–5171.

(15) Azuma, M.; Hiroi, Z.; Takano, M.; Ishida, K.; Kitaoka, Y. Observation of a Spin Gap in

SrCu$_2$O$_3$ Comprising Spin-1/2 Quasi-1D Two-Leg Ladders. *Phys. Rev. Lett.* **1994**, *73* (25), 3463–3466.

(16) Klanjšek, M.; Mayaffre, H.; Berthier, C.; Horvatić, M.; Chiari, B.; Piovesana, O.; Bouillot, P.; Kollath, C.; Orignac, E.; Citro, R.; Giamarchi, T. Controlling Luttinger liquid Physics in spin ladders under a magnetic field. *Phys. Rev. Lett.* **2008**, *101*, 137207.

(17) Tseng, Y.; Thomas, J.; Zhang, W.; Paris, E.; Pupha, P.; Bag, R.; Deng, G.; Asmara, T. C.; Strocov, V. N.; Singh, S.; Pomjakushina, E.; Kumar, U.; Nocera, A.; Rønnow, H. M.; Johnston, S.; Schmitt, T. Crossover of high-energy spin fluctuations from collective triplons to localized magnetic excitations in Sr$_{14-x}$Ca$_x$Cu$_{24}$O$_{41}$ ladders. *Npj Quantum Materials*, **2022**, 7, 92.

(18) Maeshima, N.; Hino, K. Multi-Triplon excitations of Hubbard ladders with site-dependent potentials. *J. Phys. Soc. Jpn*, **2024**, 93,054707.

(19) Horsch, P.; Sofin, M.; Mayr, M.; Jansen, M. Wigner crystallization in Na$_3$Cu$_2$O$_4$ and Na$_8$Cu$_5$O$_{10}$ chain compounds, *Phys. Rev. Lett.* **2005**, *94*, 076403.

(20) Kofu, M.; Ueda, H.; Nojiri, H.; Oshima, Y.; Zenmoto, T.; Rule, K. C.; Gerischer, S.; Lake, B.; Batista, C. D.; Ueda, Y.; Lee, S. H. Magnetic-field induced phase transitions in a weakly coupled s=1/2 quantum spin dimer system Ba$_3$Cr$_2$O$_8$, *Phys. Rev. Lett.* **2009**, *102*, 177204.

(21) Garlea, V. O.; Zheludev, A.; Masuda, T.; Manaka, H.; Regnault, L.-P.; Ressouche, E.; Grenier, B.; Chung, J.-H.; Qiu, Y.; Habicht, K.; Kiefer, K.; Boehm, M. Excitations from a Bose-Einstein Condensate of Magnons in Coupled Spin Ladders. *Phys. Rev. Lett.* **2007**, *98* (16), 167202.

(22) Kono, Y.; Kittaka, S.; Yamaguchi, H.; Hosokoshi, Y.; Sakakibara, T. Emergent critical phenomenon in spin-1/2 ferromagnetic-leg ladders : quasi-one-dimensional Bose-Einstein condensate. *Phys. Rev. B* **2019**, 100, 054442.

(23) Okamoto, K. Phase diagram of the S=1/2 two-leg spin ladder with staggered bond alternation. *Phys. Rev. B* **2003**, 67, 212408.

(24) Lorenz, T.; Heyer, O.; Garst, M.; Anfuso, F.; Rosch, A.; Rüegg, Ch.; Krämer, K. Diverging thermal expansion of the spin-Ladder system (C$_5$H$_{12}$N)$_2$CuBr$_4$. *Phys. Rev. Lett.* **2008**, *100*, 067208.

(25) Johnston, D. C.; Johnson, J. W.; Goshorn, D. P.; Jacobson, A. J. Magnetic Susceptibility of (VO)$_2$P$_2$O$_7$ : A One-Dimensional Spin- 1/2 Heisenberg Antiferromagnet with a Ladder Spin Configuration and a Singlet Ground State. *Phys. Rev. B* **1987**, *35* (1), 219–222.

(26) Pchelkina, Z. V.; Mazurenko, V. V.; Volkova, O. S.; Deeva, E. B.; Morozov, I. V.; Shutov, V. V.; Troyanov, S. I.; Werner, J.; Koo, C.; Klingeler, R.; Vasiliev, A. N. Electronic Structure and Magnetic Properties of the Strong-Rung Spin-1 Ladder Compound Rb$_3$Ni$_2$(NO$_3$)$_7$. *Phys. Rev. B* **2018**, *97* (14), 144420.

(27) Xu, Y.-J.; Zhao, H.; Chen, Y.-G.; Yan, Y.-H. Spin-Peierls Instability in the Ferromagnetic Heisenberg Ladder. *Chin. Phys. Lett.* **2013**, *30* (3), 037503.

(28) He, M.; Datta, T.; Yao, D.-X. K-edge and L$_3$-edge RIXS Study of Columnar and Staggered Quantum Dimer Phases of the Square Lattice Heisenberg Model. *Phys. Rev. B* **2020**, *101* (2), 024426.

(29) Jeong, M.; Mayaffre, H.; Berthier, C.; Schmidiger, D.; Zheludev, A.; Horvatić, M. Attractive Tomonaga-Luttinger Liquid in a Quantum Spin Ladder. *Phys. Rev. Lett.* **2013**, *111* (10), 106404.

(30) Sushkov, O. P.; Kotov, V. N. Bound States of Magnons in the S = 1/2 Quantum Spin Ladder. *Phys. Rev. Lett.* **1998**, *81* (9), 1941–1944.

(31) Schmidt, K. P.; Monien, H.; Uhrig, G. S. Rung-Singlet Phase of the S = 1/2 Two-Leg Spin-Ladder with Four-Spin Cyclic Exchange. *Phys. Rev. B* **2003**, *67* (18), 184413.

(32) Anderson, P.W. Resonating valence bonds: A new kind of insulator? *Mater. Res. Bull*. **1973**, *8*, 153.

(33) Li, Y.; Chen, G.; Tong, W.; Pi, L.; Liu, J.; Yang, Z.; Wang, X.; Zhang, Q. Rare-earth triangular lattice spin liquid: a single-crystal study of $YbMgGaO_4$. *Phys. Rev. Lett*. **2015**, *115*, 167203.

(34) Wen, J.; Yu, S.L. Li, S.; Yu, W.; Li, J. X. Experimental identification of quantum spin liquids. *NPJ. Quantum Mater.* **2019**, 4, 12.

(35) Ashtar, M.; Marwat, M. A.; Gao, Y. X.; Zhang, Z. T.; Pi, L.; Yuan, S. L.; Tian, Z. M. $REZnAl_{11}O_{19}$ (RE = Pr, Nd, Sm–Tb): A New Family of Ideal 2D Triangular Lattice Frustrated Magnets. *J. Mater. Chem. C* **2019**, *7* (32), 10073–10081.

(36) Ashtar, M.; Guo, J.; Wan, Z.; Wang, Y.; Gong, G.; Liu, Y.; Su, Y.; Tian, Z. A New Family of Disorder-Free Rare-Earth-Based Kagome Lattice Magnets: Structure and Magnetic Characterizations of $RE_3BWO_9$ (RE = Pr, Nd, Gd–Ho) Boratotungstates. *Inorgan. Chem.* **2020**, *59* (8), 5368–5376.

(37) Sanders, M. B.; Krizan, J. W.; Cava, R. J. $RE_3Sb_3Zn_2O_{14}$ (RE = La, Nd, Sm, Eu, Gd): a new family of pyrochlore derivatives with rare earth ions on a 2D Kagomé lattice. *J. Mater. Chem. C.* **2016**, *4*, 541.

(38) Zhang, Z.; Cai, Y.; Kang, J.; Ouyang, Z.; Zhang, Z.; Zhang, A.; Ji, J.; Jin, F.; Zhang, Q. Anisotropic exchange coupling and ground state phase diagram of Kitaev compound YbOCl. *Phys. Rev. Res.* **2022**, 4, 033006.

(39) Liu, A.; Song, F.; Bu, H.; Li, Z.; Ashtar, M.; Qin, Y.; Liu, D.; Xia, Z.; Li, J.; Zhang, Z.; Tong, W.; Guo, H.; Tian Z. $Ba_9RE_2(SiO_4)_6$ (RE = Ho−Yb): A Family of Rare-Earth-Based Honeycomb-Lattice Magnets. *Inorgan. Chem*. **2023,** *62*, 13867−13876.

(40) Mori, T.; Tanaka, T. Magnetic Properties of Terbium $B_{12}$ Icosahedral Boron-Rich Compounds. *J. Phys. Soc. Jpn.* **1999**, *68*, 2033–2039.

(41) Mori, T. Electron-spin-resonance study of gadolinium borosilicide: A rare-earth ladder compound, *J. Appl. Phys. 2006*, 99, 1363–1364.

(42) Malkin, B. Z.; Bud'ko, S. L.; Novikov, V.V.; Crystal-field approach to rare-earth higher borides: Dimerization, thermal, and magnetic properties of $RB_{50}$(R=Tb,Dy,Ho,Er,Tm). *Phys. Rev. Materials*. **2020**, 4, 054409.

(43) Lipina, O. A.; Surat, L. L.; Chufarov, A. Yu.; Tyutyunnik, A. P.; Enyashin, A. N.; Baklanova, Y. V.; Chvanova, A. V.; Mironov, L. Yu.; Belova, K. G.; Zubkov, V. G. Structural and Spectroscopic Characterization of a New Series of Ba$_2$RE$_2$Ge$_4$O$_{13}$ (RE = Pr, Nd, Gd, and Dy) and Ba$_2$Gd$_{2-x}$Eu$_x$Ge$_4$O$_{13}$ Tetragermanates. *Dalton Trans.* **2021**, *50* (31), 10935–10946.

(44) Wierzbicka-Wieczorek, M.; Kolitsch, U.; Tillmanns, E. *Acta Crystallogr.* **2010**, C66, i29..

(45) Toby, B. H. *EXPGUI* , a Graphical User Interface for *GSAS*. *J. Appl. Crystallogr.* **2001**, *34* (2), 210–213.

(46) Landee, C. P.; Turnbull, M. M. Review: A Gentle introduction to magnetism: Units, Fields, Theory, and Experiment. J. Coord. Chem. 2014,67(3), 375-439.

(47) Hertog, B. C.; Gingras, M. J. P. Dipolar interactions and origin of spin ice in Ising pyrochlore magnets. *Phys. Rev. Lett*. **2000**, 84, 3430-3433.

(48) Ashtar, M.; Gao, Y. X.; Wang, C. L.; Qiu, Y.; Tong, W.; Zou, Y. M.; Zhang, X. W.; Marwat, M. A.; Yuan, S. L.; Tian, Z. M. Synthesis, Structure and Magnetic Properties of Rare-Earth REMgAl$_{11}$O$_{19}$ (RE = Pr, Nd) Compounds with Two-Dimensional Triangular Lattice. *J. Alloys Compd.* **2019**, *802*, 146–151.

(49) Bu, H.; Ashtar, M.; Shiroka, T.; Walker, H. C.; Fu, Z.; Zhao, J.; Gardner, J. S.; Chen, G.; Tian, Z.; Guo, H. Gapless Triangular-Lattice Spin-Liquid Candidate PrZnAl$_{11}$O$_{19}$. *Phys. Rev. B* **2022**, *106* (13), 134428.

(50) Jensen, J.; Mackintosh, A. R. *Rare earth magnetism*, Clarendon press: Oxford, 1991.

(51) Raju, N. P.; Dion, M.; Gingras, M. J. P.; Mason, T. E.; Greedan, J. E.; Transition to long-range magnetic order in the highly frustrated insulating pyrochlore antiferromagnet Gd$_2$Ti$_2$O$_7$, *Phys. Rev. B* **1999**, *59*, 14489.

(52) Zhou, H.D.; Cheng, J. G.; Hallas, A. M.; Wiebe, C. R.; G. Li, balicas, L. Zhou, J.S.; Goodenough, J. B.; Gardner, J. S.; Choi, E. S. C. Chemical pressure effects on pyrochlore spin ice. *Phys. Rev. Lett*. **2012**, 108, 207206.

(53) Ramirez, A. P. Strongly geometrically frustrated magnets. *Annu. Rev. Mater. Sci.* **1994**, *24*, 453-480.

(54) Rinehart, J. D.; Fang, M.; Evans, W. J.; Long, J. R. Strong Exchange and Magnetic Blocking in N$_2$$^{3-}$Radical-Bridged Lanthanide Complexes, Nat. Chem. **2011**, 3, 538–542.

(55) Lin, P.-H.; Burchell, T. J.; Clérac, R.; Murugesu, M. Dinuclear dysprosium(III) single-molecule magnets with a large anisotropic barrier, Angew. Chem. Int. Ed. **2008**, 47, 8848-8851.

(56) Mori, F.; Nyui, T.; Ishida, T.; Nogami, T.; Choi, K.-Y.; Nojiri, H. Oximate-bridged trinuclear Dy-Cu-Dy complex behaving as a single-molecule magnet and its mechanistic investigation, J. Am. Chem. Soc. **2006**, 128, 1440-1441.

(57) Yaraskavitch, L. R.; Revell, H. M.; Meng, S.; Ross, K. A.; Noad, H. M. L.; Dabkowska, H. A.; Gaulin, B. D.; Kycia, J. B. Spin dynamics in the frozen state of the dipolar spin ice material Dy$_2$Ti$_2$O$_7$, Phys. Rev. B **2012**, 85, 020410(R).

(58) Mennerich, C.; Klauss, H. H.; Broekelmann, M.; Litterst, F. J.; Golze, C.; Klingerler, R.; Kataev, V.; Büchner, B.; Grossjohann, S. N.; Brenig, W.; Goiran, M.; Rakoto, H.; Broto, J. M.; Kataeva, O.; Price, D. J. Antiferromagnetic dimers of Ni(II) in the S=1 spin-ladder Na$_2$Ni$_2$(C$_2$O$_4$)$_3$(H$_2$O)$_2$. *Phys. Rev. B* **2006**, 73, 174415.

(59) Singh, Y.; Johnston, D.C. Singlet ground state in the spin-1 2 dimer compound Sr$_3$Cr$_2$O$_8$,

*Phys. Rev B* **2007**, 76, 012407.

(60) Biswal, P.; Guchhait, S.; Ghosh, S.; Sarangi, S. N.; Samal, D.; Swain, D.; Kumar, M.; Nath, R.; Crystal structure and magnetic properties of the spin-1/2 frustrated two-leg ladder compounds ($C_4H_{14}N_2$)$Cu_2X_6$ (X=Cl and Br), *Phys. Rev B* **2023**, 108,134420.

(61) Blanc, N.; Trinh, J.; Dong, L.; Bai, X.; Aczel, A. A.; Mourigal, M.; Balents, L.; Siegrist, T.; Ramirez, A. P. Quantum criticality among entangled spin chains, *Nat. Phys.* **2018**, 14, 273-276.

(62) Park, S.; Choi, Y. J.; Zhang, C. L.; Cheong, S.-W. Ferroelectricity in an S=1/2 Chain Cuprate, *Phys. Rev. Lett.* **2007**, 98, 057601.

(63) Mohanty, S.; Islam, S. S.; Winterhalter-Stocker, N.; Jesche, A.; Simutis, G.; Wang, Ch.; Guguchia, Z.; Sichelschmidt, J.; Baenitz, M.; Tsirlin, A. A.; Gegenwart, P.; Nath, R. Disordered ground state in the spin-orbit coupled $J_{eff}$ = 1/2 distorted honeycomb magnet $BiYbGeO_5$, *Phys. Rev. B* **2023**, 108, 134408.

(64) Wright, J. C.; Moos, H. W.; Colwell J. H.; Mangum, B. W.; Thornton, D. D. $DyPO_4$: a three-dimensional Ising antiferromagnet. *Phys. Rev. B* **1971**, 3, 843-858.

(65) Filippi, J.; Lasjaunias, J. C.; Ravex, A.; Tchéou, F.; Rossat-Mignod, J. Specific heat of dysprosium gallium garnet between 37 mK and 2K, Solid. Stat. Comm. *1977*, 23, 613-616.

(66) Rutherford, A.; Xing, C.; Zhou, H.; Huang, Q.; Choi, E. S.; Calder, S.; Magnetic properties of $RE_2O_2CO_3$ (RE=Pr, Nd, Gd, Tb, Dy, Ho, Er, Yb) with a rare earth-bilayer of triangular lattice. arXiv: 2407.08606v1

(67) Catanese, C. A.; Meissner, H. E. Magnetic ordering in $Dy(OH)_3$ and $Ho(OH)_3$, *Phys. Rev B* **1973**, 8, 2060-2074.

(68) Taniguchi, T.; Kawaji, Y.; Ozawa, T. C.; Nagata, Y.; Noro, Y.; Samata, H.; Lan, M. D. Antiferromagnetism of $R_2BaPdO_5$ (R=La, Nd, Pr, Sm, Eu, Gd, Dy, Ho). *J. Alloy. Compounds,* **2005**, 386, 63-69.

(69) Khatsko, E. N.; Zheludev, A.; Tranquada, J. M.; Klooster, W. T.; Knigavko, A. M.; Srivastava, R. C. Neutron scattering study of the layered Ising magnet $CsDy(MoO_4)_2$. *Low Temp. Phys.* **2004**, 30,133-139.

(70) Borowiec, M. T.; Dyakonov, V. P.; Jedrzejcazk, A.; Markovich, V. I.; Pavlyuk, A. A.; Szymczak, H.; Zubov, E. E.; Zaleski, M.; Magnetic ordering of $Dy^{3+}$ ions in $RbDy(WO_4)_2$ single crystal. *J. Low Temp. Phys.* **1998**, 110, 1003-1011.

(71) Mukherjee, P.; Wu, Y.; Lampronti, G. I.; Dutton, S. E.; Magnetic properties of monoclinic lanthanide orthoborates, $LnBO_3$, Ln=Gd,Tb,Dy, Ho, Er,Yb. *Materials. Research Bulletin*, **2018**, 98, 173-179.

(72) Lagendijk, E.; Huiskamp, W.J. Caloric and magnetic properties of two compounds having predominantly magnetic dipole-dipole interactions: $DyCl_3·6H_2O$ and $ErCl_3·6H_2O$. *Physica*, **1973**, 65(1), 118-155.

(73) Dun, Z. L.; Trinh, J.; Li, K.; Lee, M.; Chen, K. W.; Baumbach, R.; Hu, Y. F.; Wang, Y. X.; Choi, E. S.; Shastry, B. S.; Ramirez, A. P.; Zhou, H. D. Magnetic ground states of the Rare-earth tripod Kagome lattice $Mg_2RE_3Sb_3O_{14}$ (RE= Gd,Dy,Er). *Phys. Rev. Lett.* **2016**, 116, 157201.

(74) Wang, L.; Oyang, Z. W.; Li, Z. R.; Cao, J. J.; Xia, Z. C. Large magnetocaloric effect in $Gd_2Si_2O_7$ and plateau-like magnetic entropy change in $Dy_2Si_2O_7$. *J. Alloy. Compounds*. **2023**, 969, 172402.

(75) Mukherjee, P.; Hamiton, A. C. S.; Glass. H. F. J.; Dutton, S. E. Sensitivity of magnetic properties to chemical pressure in lanthanide garnet $Ln_3A_2X_3O_{12}$, Ln=Gd,Tb,Dy,Ho, A=Ca,Sc,In,Te, X=Ga,Al,Li. *J. Phys.: Condens. Matter.* **2017**, 29,405808.

(76) Cooke, A. H.; Ellis, C. J.; Gehring, K. A.; Leask, M. J. M.; Martin, D. M.; Wanklyn, B. M.; Wells, M. R.; White, R. L. Observation of a magnetically controllable Jahn Teller distortion in Dysprosium Vanadate at low temperatures. *Solid. Stat. Comm*. **1970**, 8, 689-692.

(77) Holmes, L. M.; Van Uitert, L. G.; Hecker, R. R.; Hull, G. W. Magnetic behavior of metamagnetic $DyAlO_3$. *Phys. Rev. B* **1972**, 5, 138-146.

(78) Ke, X.; Adamo, C.; Schlom, D. G.; Bernhagen, M.; Uecker, R.; Schiffer, R. Low temperature magnetism in the perovskite substrate $DyScO_3$. *Appl. Phys. Lett.* **2009**, 94, 152503.

(79) Paddison, J. A. M.; Ong, H. S.; Hamp, J. O.; Mukherjee, P.; Bai, X. J.; Tucker, M. G.; Butch, N. P.; Castelnovo, C.; Mourigal, M.; Dutton, S. E. Emergent order in the kagome Ising magnet $Dy_3Mg_2Sb_3O_{14}$. *Nat. Commun*. **2016**, 7, 13842.

(80) Ozawa, T. C.; Ikoshi, A.; Taniguchi, T.; Mizusaki, S.; Nagata, Y.; Noro, Y.; Samata, H. Low temperature magnetic properties of layered compounds: $NaLnTiO_4$ (Ln=Sm,Eu, Gd, Tb, Dy, Ho and Er). *J. Alloy. Compounds*. **2008**, 448, 38-43.

(81) Xu, X.; Won, C.; Cheong, S. W. Frustrated magnetism and ferroelectricity in a $Dy^{3+}$-based triangular lattice. *Crystals*. **2023**, 13(6) 971.

(82) Slavin, V. V.; Zvyagin, A. A.; Zvyagina, G. A.; Piryatinskaya, V. G. Magnetic field effect on the electric permittivity of rare-earth aluminium borates. *Low. Temp. Phys.* **2024**, 50, 481-488.

(83) Zhou, H. D.; Bramwell, S. T.; Cheng, J. G.; Wiebe, C. R.; Li, G.; Balicas, L.; Bloxsom, J. A.; Silverstein, H. J.; Zhou, J. S.; Goodenough, J. B.; Gardner, J. S. High pressure route to generate magnetic monopole dimers in spin ice. *Nat. Commun.* **2011**, 2, 478.

(84) Cashion, J. D.; Cooke, A. H.; Leask, J. M.; Thorp, T. L.; Wells, M. R. Crystal Growth and magnetic susceptibility of some Rare-earth compounds. Part 2 Magnetic susceptibility measurements on number of Rare-Earth compounds. *J. Materials Science*, **1968**, 3, 402-407.